\documentclass[fleqn,twoside]{article}
\usepackage{espcrc2}

\usepackage{graphicx}
\usepackage[figuresright]{rotating}

\newcommand{\pt}{\ensuremath{p_{\mathrm{T}}}}
\newcommand{\MeV}{\,\ensuremath{\mathrm{Me\kern-0.1em V}}}
\newcommand{\GeV}{\,\ensuremath{\mathrm{Ge\kern-0.1em V}}}

\newcommand{\AmS}{{\protect\the\textfont2
  A\kern-.1667em\lower.5ex\hbox{M}\kern-.125emS}}

\title{Track Reconstruction Performance in CMS}

\author{Paolo Azzurri
\thanks{on behalf of the CMS Collaboration}
\address{Scuola Normale Superiore, Piazza dei Cavalieri 7,  56126 Pisa, Italy }%
}

\begin{document}

\begin{abstract}
The expected performance of track reconstruction with LHC events using the CMS silicon tracker is presented.
Track finding and fitting is accomplished with Kalman Filter techniques that achieve efficiencies above 99\%
on single muons with $\pt >$1\GeV/c.
Difficulties arise in the context of standard LHC events with a high density of charged particles, where 
the rate of fake combinatorial tracks is very large for low \pt\  tracks, and 
nuclear interactions in the tracker material reduce the tracking efficiency for charged hadrons.
Recent improvements with the CMS track reconstruction now allow to efficiently reconstruct charged 
tracks with \pt\ down to few hundred \MeV/c and as few as three crossed layers, with a very small
fake fraction, by making use of 
an optimal rejection of fake tracks in conjunction with an iterative tracking procedure.

\vspace{1pc}
\end{abstract}

\maketitle
\section{Introduction\label{sec:intro}}
Tracking at the LHC is an experimental challenge. At the design luminosity 
of 10$^{34}$~cm$^{-2}$s$^{-1}$ the proton-proton collisions at 
$\sqrt{s}$=14~TeV will produce on average 20 superimposed events
at a bunch crossing rate of 40~MHz.
Each bunch crossing will produce on average about 2000 charged tracks in the 
$|\eta|<2.5$ range,
resulting in a density of  2.5 charged tracks per cm$^2$ at $\eta=0$,
at a distance of 4~cm
from the interaction region.

The CMS experiment~\cite{ptdr} is one of the two general purpose experiments at the LHC.
In CMS track reconstruction relies on a silicon pixel and micro-strip tracker 
immersed in a 3.8~T solenoidal magnetic field along the beam line.
The CMS reference system has origin at the detector and interaction region center,
with the $z$-axis along the beam line, the $y$-axis upwards and 
the $x$-axis in the direction of the LHC center. 
The transverse plane is denoted as the $xy$ or $r\phi$ plane. 

\section{The CMS tracker\label{sec:cmst}}
\begin{figure*}[hbt]
\vspace{-.5cm}
\includegraphics{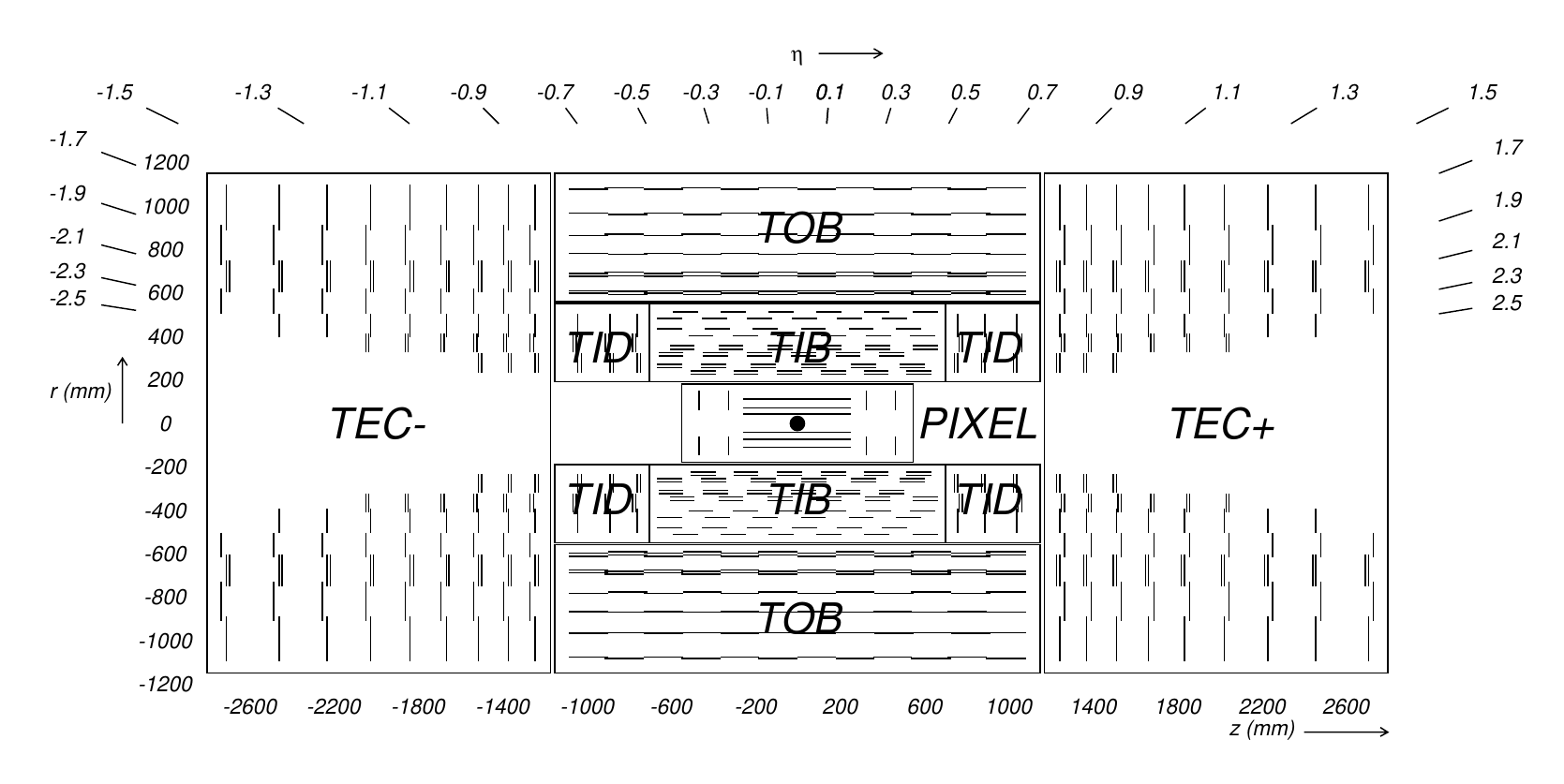}
\caption{View in the $r-z$ plane of the CMS inner tracker showing the dimensions and the 
pseudorapidity coverage. Segments represent detector modules.}
\label{fig:cmst}
\end{figure*}

The CMS inner tracker~\cite{cmst} is the largest silicon 
tracker ever built. It consist of over 15 thousand pixel and silicon strip 
modules covering an area of more than 200~m$^2$, arranged around 
the interaction point as depicted in Figure~\ref{fig:cmst}.

The pixel detector consist of about 66 million 100$\times$150~$\mu$m$^2$ pixels 
arranged in three barrel layers at 4.4, 7.3 and 10.2~cm from the beam line, 
and two endcap layers at 34.5 and 46.5~cm on each side of the interaction point.

The tracker inner barrel (TIB) and outer barrel (TOB) form ten cylindrical layers 
of silicon strip detectors around the beam line at radii ranging from 25 to 108~cm. 
The tracker inner disks (TID) and end-caps (TEC) consist respectively of three and nine
concentric ring structures extending the tracker volume to $z=\pm$270~cm along the beam line, 
covering down to $|\eta|=2.5$.

The silicon strip pitches vary from 80~$\mu$m in the inner layers to 184~$\mu$m in the outer layers.
Strips are mostly arranged in the $z$ direction in the barrel layers, yielding $r\phi$ measurements,
and in the radial direction in the endcap disks, yielding  $z\phi$ measurements.

As shown in Figure~\ref{fig:cmst}, the innermost layers and rings of each subsystem are equipped 
with back-to-back sensors mounted with a stereo angle of 100~mrad between the strips. 
These stereo layers allow $z$ measurements in the barrel layers and 
$r$ measurements in the endcaps, i.e. full three-dimensional determinations of the track 
position.
   
The CMS tracker, with  
{\it i)} an excellent silicon detector spatial resolution,  
 {\it ii)} a large level arm 
 {\it iii)} a large number of layers and
 {\it iv)} a large bending magnetic field, is designed to achieve 
 an outstanding momentum resolution for high \pt\ tracks.
 



\section{Track reconstruction\label{sec:trec}}

Different algorithms are used in CMS for track reconstruction.
All methods use the reconstructed positions (hits) of the passage of charged 
particles in the silicon detectors to determine the helix trajectories of the charged tracks 
and therefore measure their directions and momenta.
The main standard algorithm designed for the reconstruction of
proton-proton collisions is the Combinatorial Track Finder (CTF).
The CTF proceeds in three stages: (1) seeding, (2) finding and (3) fitting.

In the seeding stage pairs of hits, that are compatible with the 
interaction region above a lower \pt\ limit, are considered as 
possible candidates of charged tracks. 
Pixel hits provide the best track seeding, given their three-dimensional
position information and lower occupancy. 
The seeding efficiency with pixel hits drops in the $2<|\eta|<2.5$ 
forward region where a mixed seeding of hits from pixels and 
inner strips is needed to achieve a fully efficient track finding
in the whole tracker acceptance.

The track finding stage is based on a standard Kalman Filter 
pattern recognition approach~\cite{kalmFin}. Starting with the seeded 
parameters, the track trajectory is extrapolated to the neighboring  
tracker layers and compatible hits are assigned to the track.
The Kalman Filter is a succession of alternating
prediction and filtering steps.
At each stage the Kalman Filter updates the track parameters 
with new hits, allowing for a missing (lost) hit in a layer, in case of 
detector inefficiencies. 
The updated tracks are assigned a quality and only the best ones are kept 
for further propagation. 
Possible ambiguities with tracks sharing several hits are resolved 
in favour of the best quality trajectories. 
During the extrapolation, the uncertainties of each track trajectory
in the $r\phi$ transverse plane converge to a low level for tracks 
traversing many ($\geq 5$) layers, so that the hit assignment 
becomes fast and efficient.

The final estimate of the five parameters of each track helix 
is completed in the third stage  applying again the Kalman Filter
for the trajectory fitting~\cite{kalmFit}.
Each trajectory is refitted using a least-squares fit in two stages.
A first forward fit, inside-out from the interaction region, 
removes the approximations and biases 
of the seeding and finding stages.
A second outside-in smoother fit yields the final best estimates
of the track parameters at the origin vertex.

In the central region $|\eta|<1$, the 
obtained \pt\ resolution is better than 1\% 
for tracks with $\pt\leq 10\GeV$/c. 
At higher \pt\ the momentum resolution worsens approximately 
as  $\Delta (1/\pt)\sim 0.2$~TeV$^{-1}$.
The resolution on the transverse and longitudinal 
impact parameters in the central region is 
$\Delta d_0\sim 10~\mu$m and 
$\Delta z_0\sim 20~\mu$m, for single muons of very high 
$\pt>100~\GeV$/c.
At lower \pt\ the track impact parameter resolution worsens 
approximately with an additional
$\Delta d_0\sim \Delta z_0\sim 100 \mu{\rm m}/\pt(\GeV/{\rm c})$.

The first results obtained with
cosmic data~\cite{cerati} prove 
that the CMS tracker is ready 
for collision data.

\section{Reconstruction efficiency and fake rate}
The reconstruction of single muons with the CTF algorithm
is almost fully efficient over the whole acceptance range~\cite{ptdr}.

For charged hadrons (mostly pions) the reconstruction is more problematic due 
to their nuclear interactions in the tracker material, that corresponds to
$\sim 0.4 X_0$ at $\eta\sim0$ and grows abruptly to 
$\sim 1.8 X_0$ for $|\eta|\simeq 1.5$. 
In this situation a considerable fraction of pions sustain nuclear 
interactions while crossing the tracker and only a shorter initial
part of their trajectory can be reconstructed.

The problem with the reconstruction of tracks with few hits is 
related to their combinatorial fake rate, 
i.e. to the amount of reconstructed 
tracks that are not originating from a true charged particle.

\begin{figure}[htb]
\vspace{-.5cm}
\includegraphics[width=16pc]{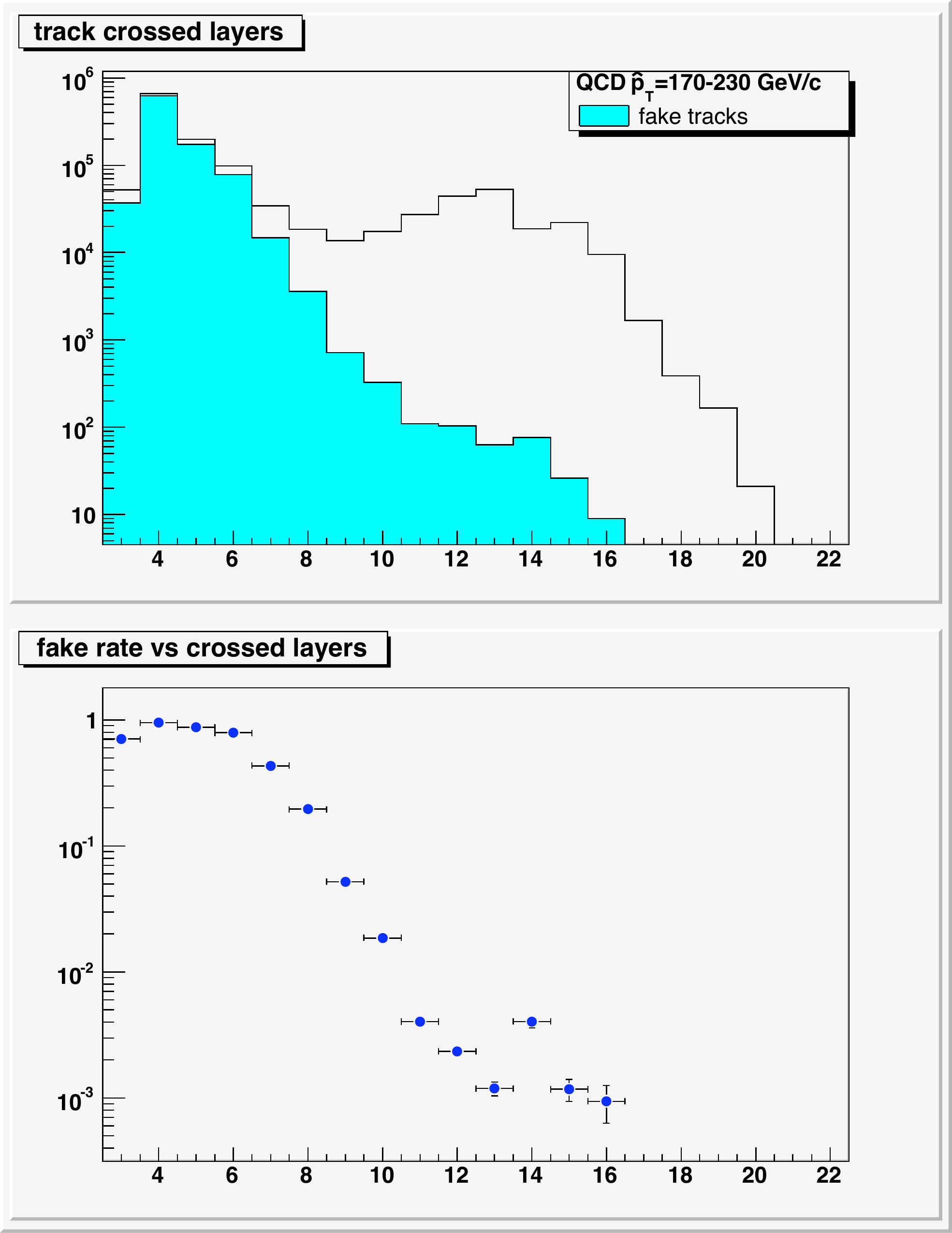}
\vspace{-.5cm}
\caption{Distribution of true and fake tracks (top) and fake rate fraction (bottom)
as a function of the number of crossed layers, using simulated QCD 
$\hat{\pt}$=170-230~GeV/c, and reconstructing tracks with ${\pt}>$300~MeV/c. }
\label{fig:fake}
\end{figure}

In a typical LHC collision event with QCD jets and a high density
of charged tracks, the CTF track finder 
yields a significant fraction of fake tracks.
The number and fraction of fake tracks as a function of the 
number of crossed layers is shown in Figure~\ref{fig:fake} 
where tracks are reconstructed with $\pt>300$~MeV/c, 
using simulated LHC collisions events  with a medium 
effective centre-of-mass
energy $\hat{\pt}$=170-230~GeV/c.

\vspace{-.5cm}

It can be seen that the number of fake tracks decreases exponentially as
a function of the number of crossed layers. The fraction of fake tracks approaches
unity for tracks crossing few layers, while tracks with many crossed layers ($>10$)
have a natural fake rate at the per mill level. 
The rate of reconstructed fake tracks also increases sensibly in the low $\pt<$1~GeV/c
region.

Analysis and results published in the CMS
Physics TDR~\cite{ptdr} used only tracks with a minimum of 8 crossed layers 
and $\pt>$0.8~GeV/c, that allowed to 
keep the fake rate at the 0.5\% level. 
This reduces the overall tracking efficiency at the ~80\% level
for tracks with $\pt>$0.9~GeV/c~\cite{ptdr},
with no possibility to reconstruct the high multiplicity of tracks $\pt<0.8$~GeV/c 
that don't even reach the CMS calorimeters. 

\section{Fake tracks filtering}
The ability to reconstruct low \pt\ and short charged tracks, with a low fake 
rate, has been recently recovered by applying a track quality filter adapted to 
the track kind. 
The rejection of fake tracks is obtained by applying optimal cuts to five 
independent variables: 1) the track fit $\chi^2$ probability, 2) the track
$d_0$ distance to the beam line and 3) measured error $\delta d_0$, 
4) the track longitudinal compatibility with the interaction vertices $\Delta z_i$
and 5) measured error $\delta \Delta z_i$.

The applied cuts are optimised as a function of the track $\pt$, $|\eta|$ 
and number of crossed layers
in order to achieve the best efficiency for a given fake rate level.
In practice no quality cuts are applied to tracks with many hits 
($>10$ crossed layers), while progressively harder cuts are
needed for tracks with fewer hits and lower \pt.
\begin{figure}[htb]
\vspace{-.5cm}
\includegraphics[width=16pc]{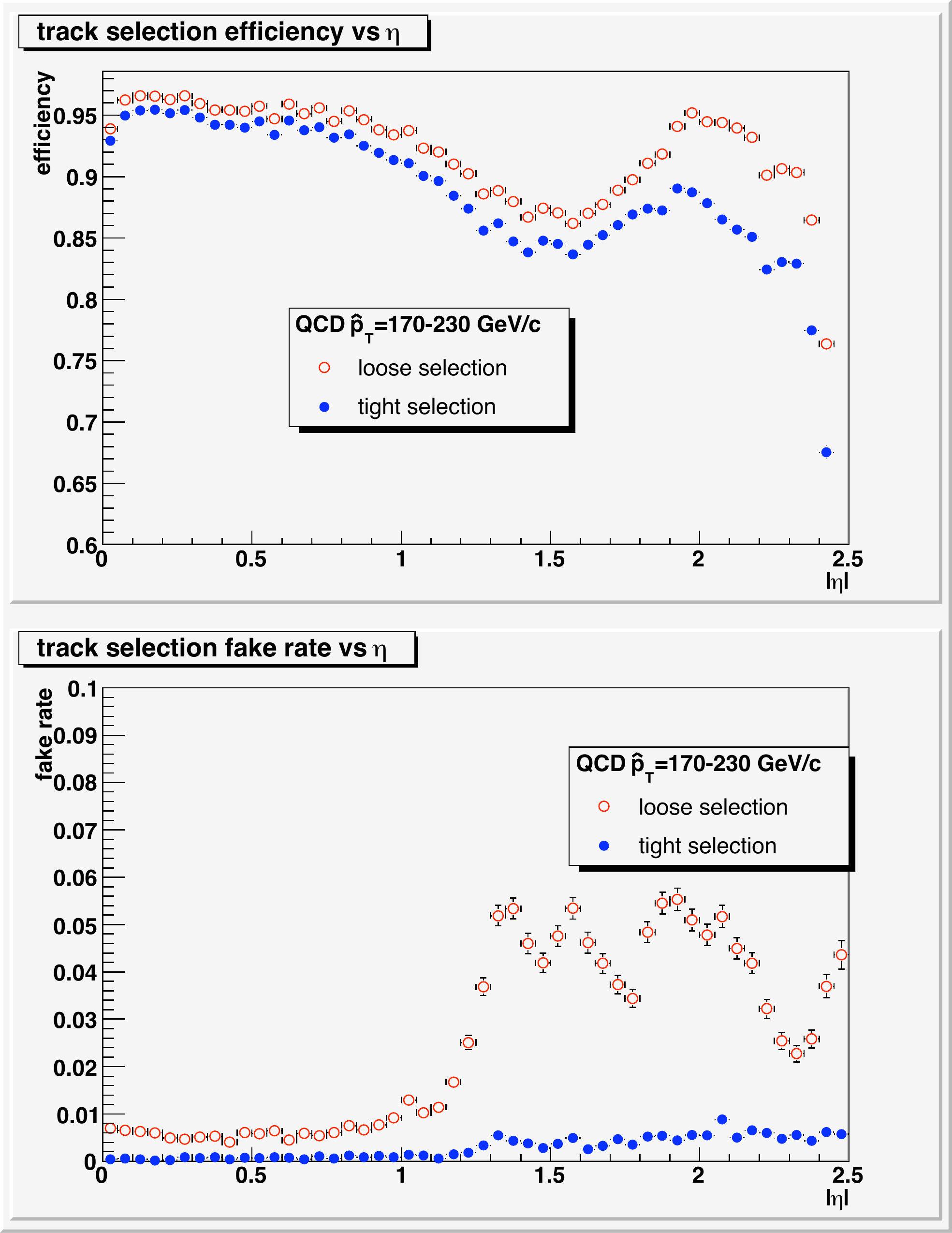}
\vspace{-.5cm}
\caption{Track reconstruction efficiency and fake rate
for all charged particles with $\pt\geq$300~\MeV/c
as a function of the track pseudorapidity $\eta$.
The looser filter selection (empty dots) ensures a higher efficiency at the 
cost of a larger fake rate in particular in the $|\eta|>1$ forward region.}
\label{fig:filter}
\end{figure}
Results are shown in Figure~\ref{fig:filter} where an average efficiency around 90\% 
for all tracks with $\pt>300$~MeV/c
is obtained with the ``tight'' selection, keeping the fake rate at the per mill level.

\section{Iterative tracking}
A further improvement to the tracking is obtained using an 
iterative procedure in the track reconstruction, i.e. running
different times the CTF algorithm. 
After a first CTF iteration a high purity filter is applied to the 
reconstructed tracks and these are put in  a first track collection. 
Hits associated with the tracks in the first collection are removed,
and the remaining hits are used for a second CTF iteration.
This procedure is repeated three or four times with a different seeding
and filtering at each iteration. 

The iterative procedure allows a faster and better reconstruction of 
charged tracks with respect to a single CTF iteration, 
raising by about 5\% the reconstruction efficiency with 
similar fake rate levels. 
This is therefore the current default 
track reconstruction procedure in CMS.

\end{document}